\documentclass[prd,numerical,
twocolumn,
showpacs]{revtex4-1}
\usepackage[english]{babel}

\usepackage{amsmath,amssymb,amsfonts,amsthm}
\usepackage{latexsym,mathrsfs}
\usepackage{color}
\usepackage{graphicx}

\definecolor{detailsgray}{gray}{0.3}

\def \rbb{\mathbb{R}}

\def \kcal {\mathcal{K}}

\def \ocal {\mathcal{O}}

\newcommand{\zetav}{\boldsymbol{\zeta}}

\newcommand{\pp}{\mathsf{p}}
\newcommand{\xx}{\mathsf{x}}
\newcommand{\yy}{\mathsf{y}}
\newcommand{\zz}{\mathsf{z}}

\def \. { \,\! }

\def\clap#1{\hbox to 0pt{\hss#1\hss}}

\newcommand{\idop}{\boldsymbol{1}}

\def \st {^\ast}

\newcommand{\Hil}{\mathcal{H}}

\newcommand{\expectation}[2]{\langle #2, #1 #2 \rangle }

\newcommand{\ad}{a^\dagger}

\newcommand{\Vin}{V_\mathrm{in}}

\newcommand{\II}{{\boldsymbol{1}}}

\newcommand{\CoinX}[1]{C_0^\infty({#1})}

\begin{document}
\title{Quantum Energy Inequality for the Massive Ising Model}
\author{Henning Bostelmann} 
\email{henning.bostelmann@york.ac.uk}
\author{Daniela Cadamuro}
\email{daniela.cadamuro@york.ac.uk}
\author{Christopher J. Fewster}
\email{chris.fewster@york.ac.uk}
\affiliation{Department of Mathematics, University of York, 
Heslington, York YO10 5DD,
United Kingdom}

\date{12 August 2015}

\begin{abstract}
A Quantum Energy Inequality (QEI) is derived for the 
massive
Ising model, giving a state-independent lower bound on suitable averages of the energy density; the first QEI to be established for an
interacting quantum field theory with nontrivial $S$-matrix. It is shown that the Ising model has one-particle states with locally negative energy densities, and that the energy density operator is not additive 
with respect to combination of one-particle states into multi-particle configurations. 
\end{abstract}
\maketitle

\section{Introduction}

In relativistic quantum field theory, the Hamiltonian is a positive
operator in all inertial frames of reference; this is the content
of the spectrum condition. By contrast, it is impossible for 
nontrivial local averages of the energy density to be positive operators in any quantum field theory obeying standard assumptions~\cite{EGJ}: typically,
the expectation value of the energy density at any point is unbounded from below with respect to the state~\cite{Fews05}.
This state of affairs demonstrates a fundamental incompatibility between
quantum fields and the energy conditions usually assumed in
classical general relativity, and which are the essential input for
results such as the singularity theorems of Penrose and Hawking~\cite{Penrose1965, Hawking1966:i, HawPen1970}, the
positive mass theorems~\cite{SchoenYau:1979,SchoenYau:1981,Witten:1981,LudvigsenVickers:1982}, and Hawking's chronology protection results~\cite{Hawking:1992},
among many others. 

Nonetheless, various models of quantum field theory obey local
remnants of the spectrum condition called Quantum Energy Inequalities (QEIs), which provide lower bounds on expectation values
of the energy density when averaged along a timelike curve or over
a spacetime region. Their study originates from Ford's insight~\cite{Ford78} that  quantum field theory could produce observable deviations from the second law of thermodynamics unless there were mechanisms to constrain negative energy densities (or fluxes). QEI bounds
place severe constraints on the extent to which quantum fields can 
support exotic spacetime geometries~\cite{FordRoman_worm, PfenningFord_warp}; moreover, weakened 
classical energy conditions inspired by QEIs can be used to
prove singularity theorems~\cite{FewsterGalloway:2011}. 

QEIs have been established for free (minimally coupled) Klein--Gordon~\cite{Ford:1991,Flanagan:1997,FordRoman:1997, PfenningFord_static:1998,FewsterEveson:1998,FewsterTeo:1999,Fews00}, Dirac~\cite{FewsterVerch_Dirac,FewsterMistry:2003,Dawson:2006,DawsFews:2006}, Maxwell~\cite{Pfenning:2001,Few&Pfen03} and Proca fields~\cite{Few&Pfen03} in both flat and curved spacetimes, for the Rarita--Schwinger field in Minkowski space~\cite{YuWu:2004},
and for the whole class of unitary positive-energy
conformal field theories in two-dimensional Minkowski space~\cite{FewsterHollands:2005} (generalising a special case~\cite{Flanagan:1997}).
In all these cases, the energy density is
 bounded from below on the class of physically acceptable
states, if it is smeared against a positive test function over a region or curve of nonzero temporal extent. Even within the setting of free fields,
however, the nonminimally coupled Klein--Gordon field provides an
example where only a weaker type of QEI holds~\cite{FewsterOsterbrink2008} -- the lower bound
is no longer state-independent, but exhibits dependence on the energy
scale -- and it is expected that this behaviour would be typical
for interacting quantum field theories~\cite{OlumGraham03}.
Analogues of these energy-dependent QEIs exist in general quantum field theories for observables arising in operator product expansions
of `classically positive' expressions~\cite{BostelmannFewster:2009}; however, a
direct connection to the energy density is lacking.

In short, QEIs are known to hold in interaction-free situations and for fields interacting with a gravitational background, but their status remains open in models with self-interaction in the sense of a nontrivial scattering matrix. This is unsatisfactory, since self-interaction is clearly expected to influence the energy density of a physical system. However, QEIs are an inherently nonperturbative concept, and the rigorous nonperturbative description of interacting quantum field theories remains challenging: it is still out of reach in physical space-time; and even in simplified low-dimensional models that are under full mathematical control, such as $P(\phi)_2$ \cite{GliJaf:quantum_physics} or integrable models \cite{Lechner:2008}, the local observables -- including the energy density -- are of considerable complexity.

We will bypass this problem here by restricting ourselves to the very simplest interacting example of an integrable quantum field theory: the massive Ising model, which has a two-particle $S$-matrix of $S_2 =-\idop$. Our object is to investigate the phenomenon of negative energy density in this model. Despite its simple scattering theory, we find that the Ising model shows clear signs of interaction: its energy density differs in essential ways from that of the free scalar field. As we will show (Sec.~\ref{sec:states}), there are single-particle states of the Ising model that have locally negative energy density, and when passing to multi-particle states, the energy density is not additive. Yet a state-independent QEI bound holds (Sec.~\ref{sec:qei}), and is given as the ($-$) case of Eq.~\eqref{eq:QEI} below; the ($+$) case applies to the free scalar field and is given for comparison.

The simplicity of the Ising model will allow us to obtain our results in a surprisingly short and elegant manner. Specifically, although it is a theory of interacting bosons, the Ising model has a free Fermi (Majorana) field closely associated with it (Sec.~\ref{sec:model}). This will allow us to adapt arguments originally developed in the context of the free Dirac field~\cite{FewsterVerch_Dirac, FewsterMistry:2003,Dawson:2006} in order to derive our QEI.

\section{The Ising model}\label{sec:model}

In this paper, we understand the Ising model of mass $\mu>0$ as a quantum field theory on 1+1 dimensional Minkowski space.  
This is usually derived as the continuum limit of a two-dimensional Ising lattice spin system above the critical temperature, which yields a (not conformally invariant) statistical theory on two-dimensional Euclidean space, from which a quantum field theory on 1+1 dimensional Minkowski space can be defined by analytic continuation of the $n$-point functions \cite{McCoyTracyWu:1977}. We will not follow this route explicitly here, but define the model directly on Minkowski space in terms of wedge-local interacting fields, in the spirit of \cite{Lechner:2008}.

Let us recall the mathematical setting of the Ising model, in a way that makes parallels to the 1+1 dimensional massive scalar free field clear. We work on the single particle space $\kcal=L^2(\rbb,d\theta)$, where the rapidity $\theta$ is related to two-momentum $\pp$ by $\pp(\theta)=\mu(\cosh \theta,\sinh\theta)$. The two models (with ($+$) for the free field and ($-$) for the Ising model) are defined on the symmetric, respectively antisymmetric, Fock space $\Hil_\pm$ over $\kcal$, with vacuum vector denoted as $\Omega$. On $\Hil_\pm$, we have the usual action of creators and annihilators  $\ad_\pm(\theta)$, $a_\pm(\eta)$, which fulfill canonical (anti-)commutation relations,
\begin{equation*}
  a_\pm (\eta)\ad_\pm(\theta) \mp \ad_\pm (\theta) a_\pm(\eta) = \delta(\theta-\eta)\idop.
\end{equation*}
Spacetime symmetries, i.e., translations $\xx=(t,x)$, boosts $\lambda$, and the space-time reflection $j$, act on $\Hil_\pm$ by
\begin{align} \label{eq:poincare}
  &U_\pm(\xx,\lambda) \ad_\pm(\theta_1)\cdots  \ad_\pm(\theta_n) \Omega =  \notag  \\ 
  &\qquad e^{i (\pp(\theta_1)+\ldots+\pp(\theta_n))\cdot \xx}  \ad_\pm(\theta_1+\lambda) \cdots  \ad_\pm(\theta_n+\lambda) \Omega,
\\ \label{eq:pct}
  &U_\pm(j) \ad_\pm(\theta_1)\cdots  \ad_\pm(\theta_n) \Omega =  \ad_\pm(\theta_n)\cdots  \ad_\pm(\theta_1) \Omega,
\end{align}
with $U_\pm(j)$, the PCT operator,  extended antilinearly. 

We now describe the basic observables of the model. Following Lechner \cite{Lechner:2008} (though with slightly different conventions), we define quantum fields $\phi_\pm$, $\phi_\pm'$ as 
\begin{align*}
 \phi_\pm(\xx) &= \frac{1}{\sqrt{4\pi}}\int d\theta \, \Big(e^{i\pp(\theta)\cdot \xx} \ad_\pm(\theta) + e^{-i\pp(\theta)\cdot \xx} a_\pm(\theta) \Big), \notag \\
\phi_\pm'(\xx) &= U_\pm(j) \phi_\pm(-\xx)  U_\pm(j).
\end{align*}
These are covariant under the symmetry operations \eqref{eq:poincare}, \eqref{eq:pct}, with space-time reflection exchanging $\phi_\pm$ and $\phi'_\pm$. Of course, $\phi_+=\phi_+'$ is just the usual local free scalar field. But $\phi_-$ is an interacting and nonlocal field; rather than local commutation relations, we have $[\phi_-(\xx),\phi'_-(\yy)]=0$ if $(\xx-\yy)^2<0$ and $\xx^1<\yy^1$ (we say $\xx$ is \emph{to the left} of $\yy$). A general operator $A$ is then considered to be localized at $\xx$ (or in a region $O$) if 
\begin{equation}\label{eq:rel_loc}
[\phi_-(\yy),A]=0=[A,\phi'_-(\zz)]
\end{equation}
whenever $\yy$ is to the left of $\xx$ (or $O$) and $\zz$ is to the 
right. By abstract arguments, a large class of such local observables exists  \cite{Lechner:2008}.

It may seem surprising, though entirely expected by general results \cite{BuchholzSummers:2006}, that this abstract notion of local observables fixes the scattering theory of the models completely \cite{Lechner:2008}. The case $\phi_+$ is trivial of course. In the case of $\phi_-$, despite its formulation on a fermionic Fock space, one finds that both the incoming and outgoing states are \emph{bosonic}, i.e., we can identify $\Hil_\mathrm{in}$ and $\Hil_\mathrm{out}$ with $\Hil_+$. The incoming M\o{}ller operator is then given by $ \Vin: \Hil_- \to \Hil_+$,
\begin{equation*}
   \Vin  \ad_-(\theta_1)\cdots \ad_-(\theta_n)\Omega =
    \Big(
     \prod_{i<j}\epsilon(\theta_i - \theta_j)
     \Big) 
     \ad_+(\theta_1)\cdots \ad_+(\theta_n)\Omega 
\end{equation*}
and the outgoing M\o{}ller operator $V_\mathrm{out}$ by a similar formula, 
but with the argument of each $\epsilon$ negated. The
$S$-matrix is $S= \Vin V_\mathrm{out}\st  = 
(-\idop)^{N_+(N_+-1)/2}$, where $N_+$ is the bosonic number operator
on $\Hil_+$. This confirms our interpretation of the case ($-$) as the interacting massive Ising model.

A crucial observation, for our purposes, is that a free Majorana field   
$\psi=(\psi_1,  \psi_2)^T$ 
can be defined  on the fermionic Fock space $\Hil_-$ by
\begin{equation*}
   \psi_{1,2}(\xx) = \sqrt{\frac{\mu}{4 \pi}} \int d\theta \,   e^{i \pi(-2\pm 1)/4}e^{i\pp(\theta)\cdot \xx\pm\theta/2} \ad_-(\theta) + 
\text{h.c.}
\end{equation*}
(here $+$ for the case $1$ and $-$ for the case $2$). 
This Majorana field is covariant under $U_-(\xx,\lambda)$ as defined above, but \emph{not} under $U_-(j)$; its associated PCT operator is fundamentally different. By analogy with \cite[Sec.~6]{BuchholzSummers:2007}, $\psi$ fulfills
\begin{equation}\label{eq:commanti}
  \{ \phi_-(\yy), \psi(\xx) \}  = 0 = [\psi(\xx),\phi_-'(\zz) ] 
\end{equation}
if $\yy$ is to the left and $\zz$ is to the right of $\xx$.
That is, $\psi(\xx)$ is \emph{not} a local observable of the interacting theory (cf.~\eqref{eq:rel_loc}). However, from \eqref{eq:commanti}, all \emph{even polynomials} in $\psi_{1,2}$ have this property. This applies in particular to the energy density of the Majorana field,
\begin{equation*}
   T^{00}_-(\xx) = \frac{i}{4} {:} \psi(\xx)^T \partial_t\psi(\xx) - (\partial_t\psi(\xx))^T \psi(\xx) {:} \, .
\end{equation*}
Since the energy-momentum operators of the Majorana field coincide with those on $\Hil_-$ by \eqref{eq:poincare}, this $T^{00}_-$ is also the energy density of the interacting Ising model. Nonetheless, we emphasize that the Ising model is distinct from the free Majorana theory; there are many local observables in the Ising model (including local fields~\cite{SchroerTruong:1978}) that do not arise from bilocal expressions in the $\psi_{1,2}$. Also, as we have mentioned, the PCT operators of the two theories are distinct. 

For completeness, we recall that the energy density of the free scalar Bose field $\phi_+$ on $\Hil_+$ is
\begin{equation*}
T_+^{00}(\xx) = \frac{1}{2}{:}(\partial_t\phi_+(\xx))^2 + (\partial_x\phi_+(\xx))^2
+\mu^2 \phi_+(\xx)^2{:}\,.
\end{equation*}
After some computation, the energy densities
of both models take the form
\begin{equation}\label{eq:T00exp}
\begin{aligned}
   T^{00}_\pm (\xx) = & \tfrac{1}{2} \smash{\int} d\theta\,d\eta \Big(  F_\pm(\theta,\eta,\xx) \ad_\pm(\theta)\ad_\pm(\eta)  
  \\ &+ 2 F_\pm(\theta,\eta+\pi i,\xx) \ad_\pm(\theta)a_\pm(\eta) \\
  &+  F_\pm(\theta+\pi i,\eta+\pi i,\xx) a_\pm(\theta) a_\pm(\eta)\Big)
\end{aligned}
\end{equation}
with
\begin{align*}
F_+(\zetav,\xx) &= -\frac{\mu^2}{2 \pi } \sinh^2 \frac{\zeta_1+\zeta_2}{2} e^{i(\pp(\zeta_1)+\pp(\zeta_2))\cdot \xx},\\
  F_-(\zetav,\xx) &= i \sinh \frac{\zeta_1-\zeta_2}{2} F_+(\zetav,\xx).	
\end{align*}
(See \cite{BostelmannCadamuro:expansion,Cadamuro:2012} for the general theory of these expansions.) Using methods from \cite[Sec.~9.1]{Cadamuro:2012}, one can see from the analyticity structure of $F_\pm$ that $T^{00}_\pm(h)=\int dt\,h(t)\,T^{00}_\pm(t,x)$ is a closable operator, and a local observable, for any $x \in \rbb$ and Schwartz test function $h$.

\section{States with negative energy density}\label{sec:states}

\begin{figure}
 \includegraphics[width=0.45\textwidth]{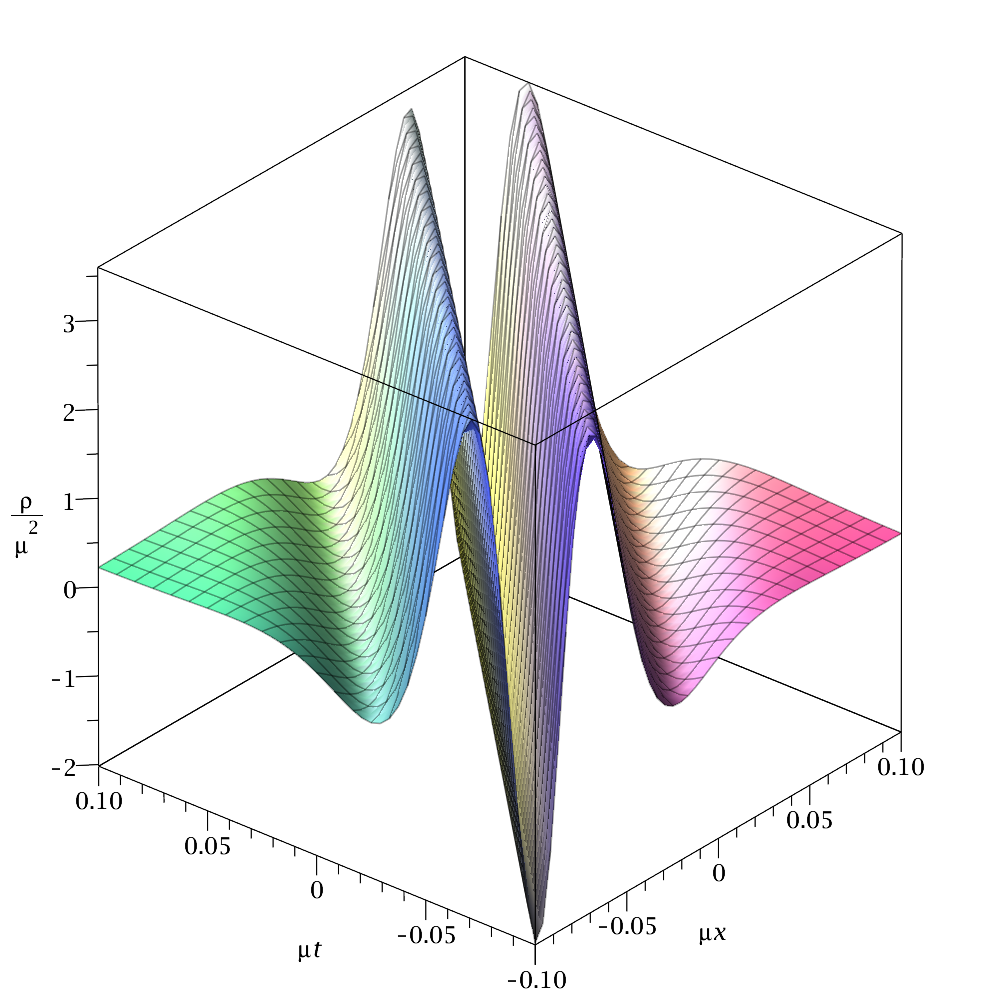}
 \caption{\label{fig:sp}(Color online) Energy density in the single-particle state $\ad_-(\varphi_{\alpha,\beta,\gamma})\Omega$ over space-time, with parameters $\alpha=0.5$, $\beta=-0.04$, $\gamma=5$, $h(\theta)=\exp(-\theta^2)/\sqrt{\pi}$.}
\end{figure}

Our first aim is to construct states with locally negative energy density, i.e., we are looking for $\Phi\in\Hil_\pm$ such that $\expectation{T^{00}_\pm(\xx)}{\Phi}<0$ for some $\xx$.

In the bosonic free field situation, it is well known (see e.g.,~\cite{Vollick:1998}) that $T^{00}_+$ has nonnegative 
expectation value in all single particle states, or more generally, states with sharp particle number, while more general superpositions (such as 
that of the vacuum and a two-particle state) can yield negative
expectation values (see, e.g.,~\cite{EGJ}). By contrast, we will now exhibit single-particle states in the Ising model with negative energy density at the origin, reminiscent of
the situation for free Dirac fields~\cite{Vollick:1998}.

These states are essentially superpositions of two plane waves. More specifically, let us choose a nonnegative real-valued Schwartz function $h$ with $\int h(\theta)  d\theta =1$. We set 
\begin{equation*}
\begin{aligned}
\Phi:= \ad_-(\varphi_{\alpha,\beta,\gamma})\Omega&=\int d\theta\, \varphi_{\alpha,\beta,\gamma}(\theta)\ad_-(\theta)\Omega,
\\
\text{where }\quad
\varphi_{\alpha,\beta,\gamma}(\theta)&:=c_{\alpha,\beta,\gamma}\big(h_{\alpha}(\theta)+\beta h_{\alpha}(\theta- \gamma)\big),
\\
h_{\alpha}(\theta)&:= \alpha^{-1}h(\alpha^{-1}\theta),
\end{aligned}
\end{equation*}
with parameters $\alpha>0$, $\beta, \gamma\in\rbb$ and 
normalization constant $c_{\alpha,\beta,\gamma}>0$. We will show that $\expectation{T^{00}_- (0)}{\Phi } <0$  for a suitable choice of the parameters. To that end, we compute from \eqref{eq:T00exp},
\begin{equation}\label{polyabc}
\expectation{T^{00}_- (0)}{\Phi } = \frac{\mu^2}{2\pi}c_{\alpha,\beta,\gamma}^2 \big(I_{\alpha} + J_{\alpha,\gamma} \beta + K_{\alpha,\gamma} \beta^2 \big),
\end{equation}
where we denoted
\begin{align*}
I_{\alpha} &= \hphantom{2} \int d\theta d\eta\, h_\alpha (\theta) h_\alpha (\eta) \cosh^2\tfrac{\theta +\eta}{2} \cosh \tfrac{\theta-\eta}{2},\\
J_{\alpha,\gamma} &= 2\int d\theta d\eta\, h_\alpha(\theta)h_\alpha(\eta) \cosh^2 \tfrac{\theta+\eta+\gamma}{2}\cosh\tfrac{\theta-\eta+\gamma}{2},\\
K_{\alpha,\gamma} &=\hphantom{2} \int d\theta d\eta\, h_\alpha (\theta) h_\alpha (\eta)\cosh^2 \tfrac{\theta +\eta+2\gamma}{2}\cosh \tfrac{\theta-\eta}{2}.
\end{align*}
The right hand side of \eqref{polyabc} is negative for some $\beta$ if the polynomial $I_{\alpha} + J_{\alpha,\gamma} \beta + K_{\alpha,\gamma} \beta^2$ has two real zeros, that is, if $J_{\alpha,\gamma} ^2 > 4 I_{\alpha} K_{\alpha,\gamma}$. This inequality holds for small $\alpha$ if it holds in the limit $\alpha \to 0$. Noting that $h_\alpha(\theta) \to \delta(\theta)$ in this limit, we obtain
\begin{equation*}
I_{0} = 1, \quad J_{0,\gamma} = 2 \cosh^3 \frac{\gamma}{2}, \quad K_{0,\gamma}=\cosh^2 \gamma.
\end{equation*}
The condition $J_{0,\gamma} ^2 > 4 I_{0} K_{0,\gamma}$ then  becomes
\begin{equation*}
(1 + \cosh\gamma)^3 > 8 \cosh^2 \gamma,
\end{equation*}
which is in fact fulfilled for sufficiently large $\gamma$. With these choices, we achieve $\expectation{T^{00}_- (0)}{\Phi } <0$. For an example see Fig.~\ref{fig:sp}, where  $\rho(t,x):=\expectation{T^{00}_- (t,x)}{\Phi }$ is plotted for a suitable choice of the parameters.

%

Let us now proceed to states with more than one particle.
Given any normalized single-particle state wave function $\varphi\in\kcal$,
we may form multi-particle states by taking tensor products in the `in'
Hilbert space and applying the inverse M\o{}ller operator. For the free model, this yields $n$-particle vectors $\Phi_{n+}:= (n!)^{-1/2} \ad_+(\varphi)^n \Omega \in\Hil_+$, 
while the corresponding states in the Ising model are $\Phi_{n-}:= \Vin\st \Phi_{n+} \in \Hil_-$.  

Now, the total energy, given by the Hamiltonian $H_\pm$ (the generator of time translations), is additive in the sense that $\expectation{H_\pm}{ \Phi_{n\pm}} = n \expectation{H_\pm}{ \Phi_{1\pm}}$. (For $H_+$ this is evident from \eqref{eq:poincare}, and for $H_-$ it follows since $\Vin H_- \Vin\st = H_+$.) Furthermore, the energy density in the free model is also additive: 
\begin{equation*}
\expectation{T^{00}_+ (\xx) }{ \Phi_{n+}} = n \expectation{T^{00}_+ (\xx)}{ \Phi_{1+}}.
\end{equation*}
However, the same relation does \emph{not} hold in the interacting situation: In general,
\begin{equation}\label{eq:nonadditive}
\expectation{T^{00}_- (\xx) }{ \Phi_{n-}} \neq n \expectation{T^{00}_- (\xx) }{\Phi_{1-}}.
\end{equation}
We can deduce this from our other results: We saw above that $\expectation{T^{00}_- (\xx) }{ \Phi_{1-}}$ is negative in some examples, and therefore at large $n$, equality in \eqref{eq:nonadditive} would be in contradiction to the QEI \eqref{eq:QEI} that we will establish below.

\begin{figure}
 \includegraphics[width=0.45\textwidth]{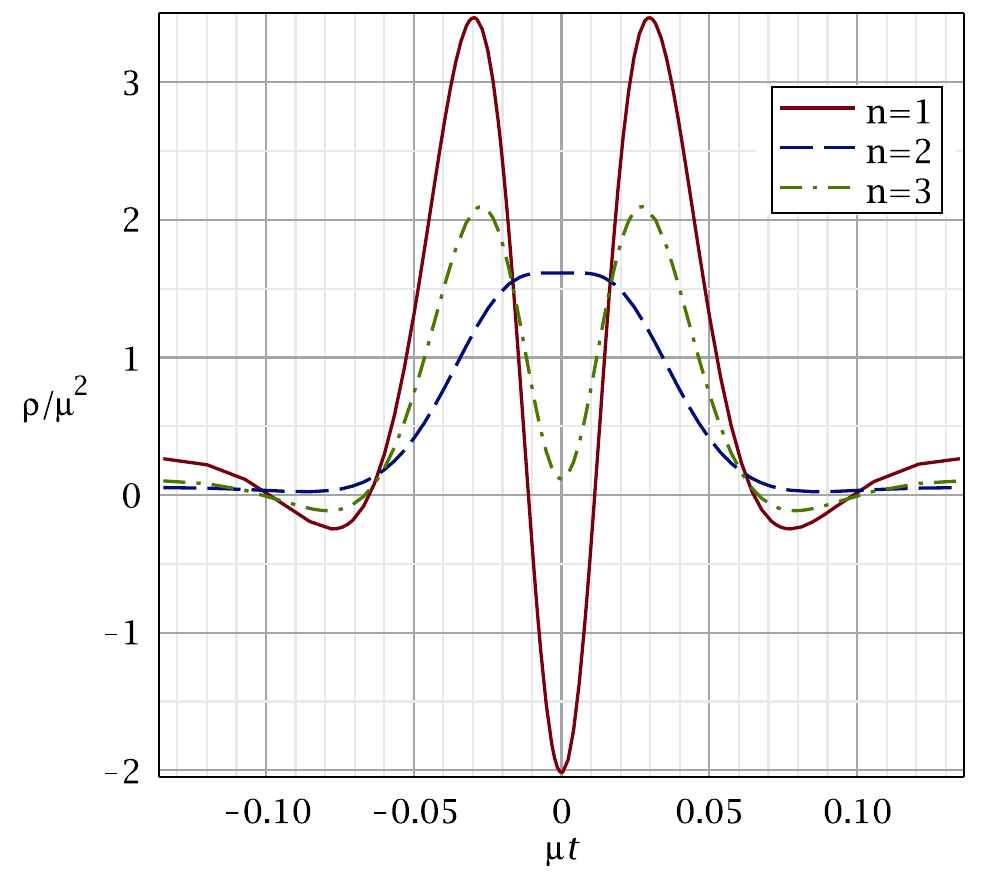}
 \caption{\label{fig:mp}(Color online) Energy density per particle in the Ising model, computed in the multi particle states $\Phi_{n-}$ along the time axis, with parameters as in Fig.~\ref{fig:sp}}
\end{figure}

But let us give a direct argument for \eqref{eq:nonadditive}. It is useful to note that $\Vin \ad_-(\theta) \Vin\st = \ad_+(\theta) M(\theta)$, where $M(\theta)$ is the multiplication operator
\begin{equation*}\label{Mepsilon}
\begin{aligned}
M(\theta) &\ad_+(\eta_1)\cdots \ad_+(\eta_n) \Omega= \\
  &\Big(\prod_{j=1}^{n}\epsilon(\theta - \eta_j)\Big) \ad_+(\eta_1)\cdots \ad_+(\eta_n) \Omega.
\end{aligned}
\end{equation*}
Using this relation and its adjoint, it is straightforward to compute from \eqref{eq:T00exp} that 
\begin{multline*}
\expectation{T^{00}_- (\xx)  }{ \Phi_{n-}} = n \, \frac{\mu^2}{2\pi} \int d\theta d\eta \cosh^2 \tfrac{\theta+\eta}{2} \cosh\tfrac{\theta-\eta}{2} \\ 
\times \overline{\varphi(\theta)} \varphi(\eta) L_{\varphi}(\theta,\eta)^{n-1} e^{i(\pp(\theta)-\pp(\eta))\cdot \xx},
\end{multline*}
where
\begin{equation*}
L_{\varphi}(\theta, \eta):=\int d\lambda \,|\varphi(\lambda)|^2  \,\epsilon(\theta - \lambda) \epsilon(\eta -\lambda).
\end{equation*}
The factor $L_\varphi$, which does not occur in the free case, prevents additivity of the energy density. 
This nonlinearity is also apparent in Fig.~\ref{fig:mp}, where the energy density per particle, $\rho(t):= \frac{1}{n} \expectation{T^{00}_- (t,0) }{\Phi_{n-}}$, is plotted along the time axis for the same choice of $\varphi=\varphi_{\alpha,\beta,\gamma}$ as above. This behaviour reflects the interaction in the Ising model.

\section{The Quantum Energy Inequality}\label{sec:qei}

Finally, we turn to the derivation of the QEI bounds on the energy density.
In fact, the free scalar field ($+$) and the Ising model ($-$) 
obey closely related QEIs,  namely
\begin{equation}\label{eq:QEI}
\smash{\int} dt\,  g(t)^2 \expectation{T_\pm^{00}(t,x)}{\Phi}  \ge -\frac{1}{4\pi^2}\smash{\int_\mu^\infty} \!\!\!\! 
d\omega\, \omega^2
|\widetilde{g}(\omega)|^2 Q_\pm \big(\frac{\omega}{\mu}\big),
\end{equation}
for any real-valued smooth compactly supported function $g$
and all sufficiently regular normalized states $\Phi$, where
the dimensionless functions $Q_\pm:[1,\infty)\to \rbb^+$
are
\begin{equation}\label{eq:Qpm}
Q_\pm(u) = \sqrt{1-u^{-2}} \pm u^{-2}\log(u+\sqrt{u^2-1})
\end{equation}
and obey $Q_\pm(1)=0$, $\lim_{u\to+\infty} Q_\pm(u)=1$,
while the Fourier transform is 
$\widetilde{g}(\omega) = \int dt\, g(t)e^{i\omega t}$.

The QEI for the free Bose field ($+$) has been known for
some time~\cite{FewsterEveson:1998} and holds rigorously
for all Hadamard states $\Phi$~\cite{Fews00}; our goal is to establish the QEI for the Ising model ($-$).  
As we have already seen that its energy
density coincides with that of the free Majorana field, the 
QEI is exactly the same as for the latter theory. This has not been
computed before, but the argument is sufficiently similar to treatments
in~\cite{FewsterMistry:2003,Dawson:2006} that we only sketch it here. 
For $\nu\in\rbb$, let $R_\nu$ and $S_\nu$ be continuous one-parameter families in $\kcal=L^2(\rbb,d\theta)$, and define for each
$\nu$ the (bounded) operator
\begin{equation*}
\ocal_\nu =\int d\theta\, \left(R_\nu(\theta)a_-(\theta) + S_\nu(\theta)a_-^\dagger(\theta)\right)
\end{equation*}
on $\Hil_-$. Elementary use of the CARs shows that
\begin{equation*}
\ocal_\nu^\dagger\ocal_\nu -\|S_\nu\|^2\II = 
-\ocal_\nu\ocal_\nu^\dagger  + \|R_\nu\|^2\II 
=:X_\nu ,
\end{equation*}
where 
$\|\cdot\|$ is the norm in $\kcal$. As  $\nu\ocal_\nu^\dagger\ocal_\nu$ (resp., $-\nu\ocal_\nu\ocal_\nu^\dagger$) is positive semidefinite for $\nu\ge 0$
(resp., $\nu\le 0$), we have
\begin{equation}\label{eq:preQI}
\int_{-\infty}^\infty \frac{d\nu}{\pi}\,\nu \expectation{X_\nu}{\Phi} \ge
-\int_{0}^\infty \frac{d\nu}{\pi}\nu \left(\|R_{-\nu}\|^2 + \|S_\nu\|^2 
\right) 
\end{equation}
for all normalized quantum states $\Phi$. For our application, we put
\begin{align*}
R_\nu(\theta) &=\sqrt{\frac{\mu}{4\pi}}  \widetilde{g}(\nu-\mu\cosh\theta) \cosh\frac{\theta}{2}, \\
S_\nu(\theta) &=- i\sqrt{\frac{\mu}{4\pi}}  \widetilde{g}(\nu+\mu\cosh\theta) \sinh\frac{\theta}{2},
\end{align*}
for any real-valued $g\in\CoinX{\rbb}$. Using 
the identity
\begin{equation*}
(\omega+\omega')\widetilde{g^2}(\omega'-\omega) =
 -\int_{-\infty}^\infty \frac{d\nu}{\pi}
\nu \overline{\tilde{g}(\nu+\omega)}\tilde{g}(\nu+\omega')
\end{equation*}
(a mild rewriting of Eq.~(2.17) in \cite{FewsterMistry:2003}), 
the left-hand side of \eqref{eq:preQI} becomes, after a computation,
\begin{equation*}
\int_{-\infty}^\infty \frac{d\nu}{\pi}\,\nu 
\expectation{X_\nu}{\Phi}
=\int dt\, g(t)^2  \expectation{T_\pm^{00}(t,0)}{\Phi}.
\end{equation*}
It is expected
that this is rigorously valid at least for all $\Phi\in\Hil_-$ that are
Hadamard states of the Majorana
field (cf.\ the treatment of the Dirac equation in four
dimensional curved spacetimes~\cite{DawsFews:2006}). 
We also compute 
\begin{align}\label{eq:QIrhs}
\text{RHS of~\eqref{eq:preQI}}& = -\frac{\mu}{4\pi^2}\int_0^\infty \!\!\!\!\!\! d\nu\,\nu \!\!\int_{-\infty}^\infty\!\!\!\!\!\! d\theta\, \cosh\theta
\lvert \tilde{g}(\mu\cosh\theta + \nu) \rvert^2  \notag\\
&= -\frac{1}{4\pi^2}\int_\mu^\infty\!\!\!\!\!\!  d\omega\,
\omega^2 \lvert \tilde{g}(\omega) \rvert^2 Q_-(\omega/\mu),
\end{align}
where $Q_-$ was defined in \eqref{eq:Qpm}.
In more detail, 
the second equality in~\eqref{eq:QIrhs} is obtained by
using evenness of the integrand to alter the inner integration
region to $[0,\infty)$, then changing variables from $(\nu,\theta)$
to $(\omega,\theta)$ where $\omega=\nu+\mu\cosh\theta$ (with a
consequent change of integration region) and 
finally evaluating the $\theta$ integral. For $x=0$,
the (common) energy density of the Majorana and Ising models thus
satisfies the QEI given as the ($-$) case of~\eqref{eq:QEI},
for all real-valued test functions $g$ and for a large domain of
$\Phi\in\Hil_-$; the result for general $x$ follows by
translation invariance. This bound
is precisely half of the bound for the free massive Dirac field
in two dimensions obtained in~\cite{Dawson:2006} using similar arguments, 
as might be expected. 

It is also worth considering the limit of these QEIs as $\mu\to 0$,
for fixed $g$, corresponding to the short-distance scaling limit of the theory \cite{BostelmannLechnerMorsella:2011}. In both cases,
\begin{equation*}
\text{RHS of~\eqref{eq:QEI}} 
\longrightarrow
-\frac{1}{4\pi}\int_{-\infty}^\infty dt\, |g'(t)|^2,
\end{equation*}
where we have used the fact that $|\tilde{g}(\omega)|$ is
even, and the Plancherel theorem. But both the massless
free scalar field and the massless Majorana field are conformal field theories, and obey sharp QEIs~\cite[Eq.~(4.25)]{FewsterHollands:2005}
\begin{equation*}
\int dt\, g(t)^2  \expectation{T_\pm^{00}(t,x)}{\Phi} \ge -\frac{C_\pm}{6\pi}\int_{-\infty}^\infty dt\, |g'(t)|^2 
\end{equation*}
for suitable normalized $\Phi$,
where $C_\pm$ are the central charges of the left- and right- moving
components: $C_+=1$ (free scalar field) and $C_-=\frac{1}{2}$ (free Majorana), so the sharp bound is therefore tighter by a factor of $3/2$,
respectively $3$, in these two cases.  Accordingly, we do not expect our QEI to be sharp for $\mu>0$.

\section{Conclusion}

In this paper, for the first time, a QEI has been derived for a quantum field theory with nontrivial $S$-matrix. Even if the $S$-matrix is rather simple in the case at hand, this underpins the expectation that QEIs are a general consistency property of relativistic quantum physics which is stable against the introduction of self-interaction.

We also saw that the interacting model allows for negative energy densities in single-particle states. In a free theory one could combine such single-particle states to 
obtain multi-particle configurations with arbitrarily large negative energy densities (cf.~\cite{FewsterOsterbrink2008}), thus excluding
the existence of a state independent QEI. However, the energy density of the Ising model is not additive with respect to tensor products
of single-particle states; a signature of the interacting nature of the theory. It is intriguing that the interaction is responsible
for maintaining the QEI in this sense.

We hope the present methods will prove to be a
foundation for similar results on other integrable theories in two spacetime dimensions, such as the sinh-Gordon model. Intriguingly, the interaction of the Ising model conspires to yield
a QEI with a state-independent lower bound; it will be interesting to see whether this persists in other models.

\begin{acknowledgments}
The authors thank Atsushi Higuchi for his perceptive questions
and comments. 
\end{acknowledgments}

\bibliography{../../aps_integrable}

\end{document}